# Twin-vortex solitons in nonlocal nonlinear media

Fangwei Ye,[1] Yaroslav V. Kartashov,[2] Bambi Hu[1,3] and Lluis Torner[2]

[1]*Department of Physics, Centre for Nonlinear Studies, and The Beijing-Hong Kong Singapore Joint Centre for Nonlinear and Complex Systems, Hong Kong Baptist University, Kowloon Tong, China*

[2]*ICFO-Institut de Ciencies Fotoniques, and Universitat Politecnica de Catalunya, Mediterranean Technology Park, 08860 Castelldefels (Barcelona), Spain*

[3]*Department of Physics, University of Houston, Houston, Texas 77204-5005, USA*

We consider soliton formation in thermal nonlinear media bounded by rectangular cross-sections and uncover a new class of nonlinear stationary topological state. Specifically, we find that stationary higher-order vortex states in standard shapes do not exist, but rather they take the form of multiple, spatially separated single-charge singularities nested in an elliptical beam. Double-charge states are found to be remarkably robust despite their shape asymmetry and the phase-singularity splitting. States with higher topological charges are found to be unstable.

*OCIS codes: 190.5530, 190.4360, 060.1810*

Light beams with nested phase-singularities, or vortices, appear in many areas of Optics. When material nonlinearity counterbalances diffraction the localized, bright vortex solitons may exist [1,2]. However, in most conservative uniform media with local focusing nonlinearity, vortex solitons are prone to azimuthal instabilities that lead to their decay [3-5]. Collapse and azimuthal instabilities can be avoided, e.g., in nonlocal nonlinear media [3,6-14], thus charge-1 vortex solitons were experimentally observed in thermal media [14]. The response of nonlocal medium can be characterized by a response function whose characteristic transverse scale determines the degree of nonlocality. In thermal media this characteristic transverse scale is determined by the size of the sample [14,15] and light propagation is strongly affected by the sample geometry and boundary conditions [16-20]. Analogous effects occur in liquid crystals [21]. In settings where boundary conditions strongly affect the nonlinear material response, the sample asymmetry may translate into significant modifications of soliton properties. In this Letter we reveal that this is the case for vortex solitons in rectangular thermal media. We find that, in contrast to previously studied cases, stationary vortex solitons with total charge $m$ always involve $m$ *spatially separated* charge-1 singularities nested in slightly elliptical beam. Remarkably, such states are not bound states of several beams carrying vorticity, but rather several spatially-separated singularities nested in a single host beam. At $m = 2$ such vortex states are stable under propagation as well as robust against modifications in sample aspect ratio. States with $m > 2$ are unstable. Note that the stationary states described here are fundamentally different from, e.g., the previously observed dynamical splitting of the dislocations nested in a high-charge dark vortex soliton propagating in photorefractive crystals with anisotropic defocusing nonlinearity [22]. In sharp contrast, the new vortex states described here exist as stationary, stable states.

A light beam propagating in a thermal nonlinear medium experiences slight absorption and acts as a heat source. The heat diffusion results in a transversely inhomogeneous temperature distribution depending on the sample geometry and on the boundary conditions. Such a temperature redistribution results in a refractive index profile that is proportional to the local temperature change. The propagation of a light beam under such conditions is described by the following system of equations

$$i\frac{\partial q}{\partial \xi} = -\frac{1}{2}\left(\frac{\partial^2 q}{\partial \eta^2} + \frac{\partial^2 q}{\partial \zeta^2}\right) - qn,$$
$$\frac{\partial^2 n}{\partial \eta^2} + \frac{\partial^2 n}{\partial \zeta^2} = -|q|^2.$$
(1)

Here $q = (k_0^2 r_0^4 \alpha\beta / \kappa n_0)^{1/2} A$ is the dimensionless light field amplitude; $n = k_0^2 r_0^2 \delta n / n_0$ is proportional to the nonlinear change $\delta n$ in the refractive index $n_0$; $\alpha$, $\beta$ and $\kappa$ are the optical absorption, thermo-optic, and thermal conductivity coefficients, respectively; the transverse $\eta, \zeta$ and longitudinal $\xi$ coordinates are scaled to the characteristic beam width $r_0$ and diffraction length $L_{\text{dif}} = k_0 r_0^2$, respectively.

The conditions imposed at the boundary of the thermal sample greatly affect the entire optically-induced refractive index profile. Here we are interested in situations where the refractive index distribution exhibits form-anisotropy induced by geometry. We thus consider the samples with rectangular cross sections elongated in one direction and assume that the boundaries of the sample with dimensions $d_\eta \times d_\zeta$ are kept at equal temperatures. We solve Eq. (1) with boundary conditions $n, q|_{\eta=0,d_\eta} = 0$ and $n, q|_{\zeta=0,d_\zeta} = 0$. We set $d_\zeta = 40$ and vary $d_\eta$. Equation (1) conserves the energy flow $U = \int\int_{-\infty}^{\infty} |q|^2 d\eta d\zeta$.

Vortex solitons of Eq. (1) are sought in the form $q(\eta,\zeta,\xi) = (w_\mathrm{r} + iw_\mathrm{i})\exp(ib\xi)$, where $b$ is the propagation constant. The topological charge of the phase singularities nested in the beam is determined by the phase variation on a closed contour surrounding singularity, i.e., $m = (2\pi)^{-1}\oint \arctan(w_\mathrm{i}/w_\mathrm{r})$. In local nonlinear media or in square thermal samples with $d_\eta = d_\zeta$, vortex solitons carry a single phase dislocation with the total topological charge $m$ nested in the center of the ring-like shape.

The central prediction of this Letter is that in thermal samples with rectangular cross-sections, stationary solutions with high topological charges $m > 1$ feature a shape where *the $m$ single-charge dislocations are spatially separated in a single host beam*. Figure 1 shows the field modulus and the phase distributions for such new class of vortex solitons, with topological charges $m = 1, 2, ..., 6$ in a sample with $d_\eta = 20$, $d_\zeta = 40$. The array of $m$ dislocations is aligned along the short axis of the sample, while intuitively one may expect that the opposite arrangement may be more favorable. Due to the multiple singularities nested in the beam core, the vortex states exhibit azimuthally modulated intensity distributions, whose maxima and minima are achieved along short and long sample axes, respectively. Therefore, usual radially-symmetric vortex solitons carrying a single multiple-charge phase dislocation do not exist in rectangular thermal samples. This may be expected taking into account that high-charge vortices nested in linear fields tend to dynamically unfold and thus split into their single-charge constituents in the presence of most asymmetric external perturbations. The remarkable result is that, under the conditions of the nonlocal medium addressed here, the resulting complex states exist as truly stationary nonlinear states.

Figures 2 describes the properties of such stationary states. For all values of $d_\eta$ and $m$ the energy flow is found to be a monotonically increasing function of $b$ [Fig. 2(a)]. In samples with sufficiently large aspect ratios $d_\zeta/d_\eta$, stationary solutions exist only above a lower threshold in terms of the propagation constant (or energy flow). Such thresholds occur also for vortices with larger number of singularities. The integral widths of the solutions along the $\eta$ and $\zeta$ axes, denoted by $W_\eta$ and $W_\zeta$, respectively, decrease with decreasing $d_\eta$, and in all cases $W_\eta \geq W_\zeta$ [Fig. 2(b)]. When the sample becomes symmetric (i.e. when $d_\eta = d_\zeta$) the integral widths $W_{\eta,\zeta}$ become equal. An increase of the ratio $d_\zeta/d_\eta$ is accompanied by a growth of the ellipticity of the overall host beam shape. At fixed $b$, $d_{\eta,\zeta}$ the ellipticity increases with vortex charge $m$.

The separation between the single-charge singularities nested in the new vortex states strongly depends on the sample aspect ratio and on the propagation constant (thus, the energy flow), as shown in Fig. 3. The distance $D_\mathrm{s}$ between two singularities for $m = 2$ solu-tions increases with decreasing $d_\eta$ [Fig. 3(a)]. $D_s$ increases most rapidly with decrease of $d_\eta$ in the interval $d_\eta \in [20, 40]$, but with further decrease of $d_\eta$ the value $D_s$ saturates. On the other hand, for a fixed sample aspect ratio, $D_s$ is a monotonically decreasing function of the propagation constant and it asymptotically approaches zero as $b \to \infty$ [Fig. 3(b)]. This is a consequence of the progressively higher localization of vortex solitons that is accompanied by the development of more symmetric refractive index distribution around the center of the sample.

Owing to the complex shapes of the high-charge stationary states described above, one may expect that all of them are dynamically unstable. To elucidate stability we performed comprehensive simulations of Eq. (1) with the input conditions $q|_{\xi=0} = (w_r + iw_i)(1+\rho)$, where $\rho$ is a small random or regular perturbation, such that $\rho(\eta, \zeta) \ll w_{r,i}(\eta, \zeta)$. As expected, fundamental vortex solitons with $m = 1$ were found always stable, for all values of the sample aspect ratio [Figs. 4(a) and 4(b)]. Remarkably, we also found that double-charge solutions, to be perhaps referred as twin-vortex solitons, are stable too despite their outstanding shapes with two well-separated singularities [Figs. 4(c) and 4(d)]. Such twin-vortex solitons were found to be stable in the entire domain of their existence. In addition, they appear to be remarkably robust against form-perturbations, as they survive modifications of the sample aspect ratio. An example is shown in Figs. 4(g)-4(i), where a twin-vortex obtained for $d_\eta = 20$, $d_\zeta = 40$ was launched into sample with $d_\eta = 21$, $d_\zeta = 39$. Despite the perturbation in the sample shape, the twin-vortex maintains its internal structure, while the line passing through two dislocations oscillates periodically, but does not rotate upon propagation. On the other hand, solutions with higher charges, i.e., $m \geq 3$, did meet with expectations and were found to be always unstable [see Figs. 4(e) and 4(f)].

We thus conclude by stressing that we reported a new class of stationary, stable, robust nonlinear vortex state in the form of a vortex-twin, featuring two identical spatially separated single-charge phase-dislocations, nested in a single host beam. Such new states were found to exist in the particular case of thermal nonlinear media bounded in rectangular cross sections. Higher-order solutions, featuring vortex arrays nested in a beam also exist, but they appear to be always unstable.

# References with titles

# References without titles

# Figure captions

Figure 1. Field modulus (top) and phase (bottom) distributions for vortex solitons in thermal medium with $d_\eta = 20$, $d_\zeta = 40$ for (a) $b = 2$, $m = 1$, (b) $b = 2$, $m = 2$, (c) $b = 4.64$, $m = 3$, (d) $b = 5$, $m = 4$, (e) $b = 5$, $m = 5$, and (f) $b = 6$, $m = 6$.

Figure 2. (a) $U$ versus $b$ for vortex with $m = 2$. (b) $W_{\eta,\zeta}$ versus $d_\eta$ for vortex with $m = 2$, $b = 2$. In all cases $d_\zeta = 40$. Circles correspond to soliton in Fig. 1(b).

Figure 3. Distance between singularities for $m = 2$ vortex (a) versus $d_\eta$ at $b = 2$, and (b) versus $b$ at $d_\eta = 20$. In all cases $d_\zeta = 40$.

Figure 4. Propagation dynamics of vortex solitons perturbed by input noise with $m = 1$, $b = 2$ (a),(b), $m = 2$, $b = 2$ (c),(d), $m = 3$, $b = 6$ (e),(f) in a sample with $d_\eta = 20$, $d_\zeta = 40$. (g)-(i) Propagation of vortex with $m = 2$, $b = 2$ that would be exact soliton in a sample with $d_\eta = 20$, $d_\zeta = 40$ in a sample with slightly changed dimensions $d_\eta = 21$, $d_\zeta = 39$.

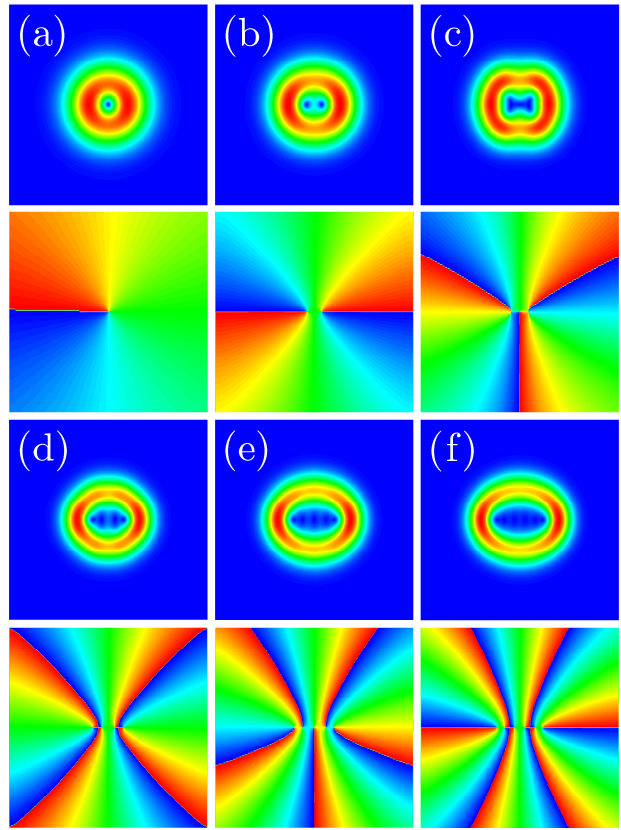

Figure 1. Field modulus (top) and phase (bottom) distributions for vortex solitons in thermal medium with $d_\eta = 20$, $d_\zeta = 40$ for (a) $b = 2$, $m = 1$, (b) $b = 2$, $m = 2$, (c) $b = 4.64$, $m = 3$, (d) $b = 5$, $m = 4$, (e) $b = 5$, $m = 5$, and (f) $b = 6$, $m = 6$.

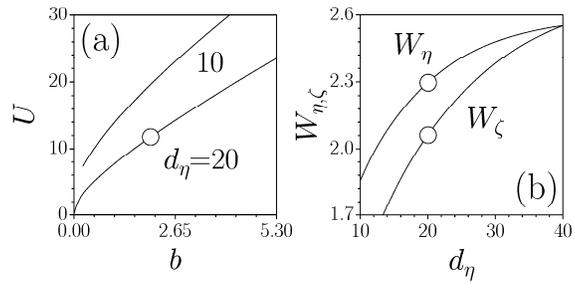

Figure 2.  (a) $U$ versus $b$ for vortex with $m = 2$. (b) $W_{\eta,\zeta}$ versus $d_\eta$ for vortex with $m = 2$, $b = 2$. In all cases $d_\zeta = 40$. Circles correspond to soliton in Fig. 1(b).

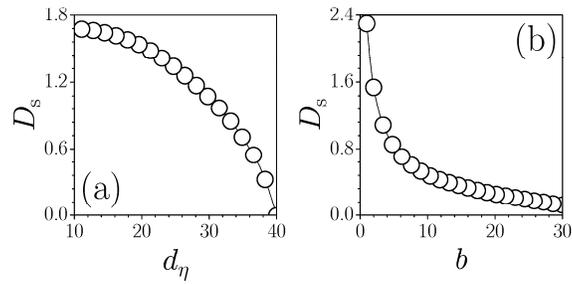

Figure 3. Distance between singularities for $m = 2$ vortex (a) versus $d_\eta$ at $b = 2$, and (b) versus $b$ at $d_\eta = 20$. In all cases $d_\zeta = 40$.

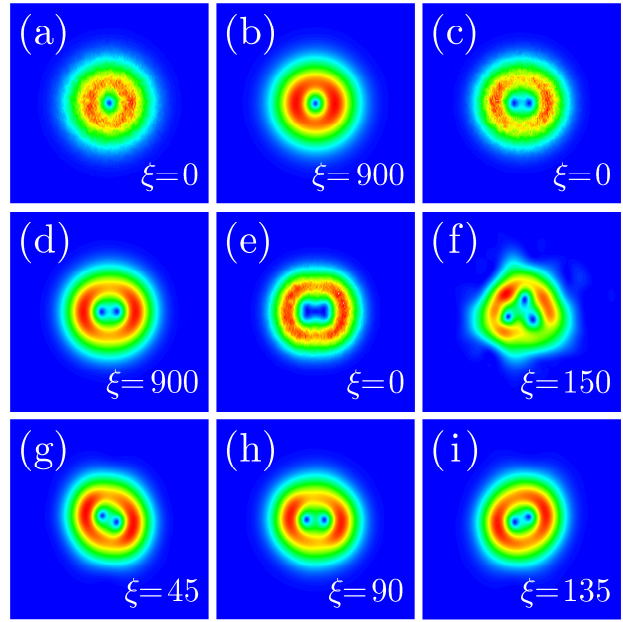

Figure 4. Propagation dynamics of vortex solitons perturbed by input noise with $m=1$, $b=2$ (a),(b), $m=2$, $b=2$ (c),(d), $m=3$, $b=6$ (e),(f) in a sample with $d_\eta=20$, $d_\zeta=40$. (g)-(i) Propagation of vortex with $m=2$, $b=2$ that would be exact soliton in a sample with $d_\eta=20$, $d_\zeta=40$ in a sample with slightly changed dimensions $d_\eta=21$, $d_\zeta=39$.